**Evaluating the Impact of State-Level Public Masking Mandates on New COVID-19 Cases and Deaths in the United States: A Demonstration of the Causal Roadmap**

**Authors:** Angus K. Wong and Laura B. Balzer

**Correspondence to:** Angus K. Wong, Department of Biostatistics & Epidemiology, University of Massachusetts Amherst, 427 Arnold House, Amherst, MA, 01003 (email: angus.ko.wong@gmail.com, phone: 1-339-223-1484)

**Author Affiliations:** Department of Biostatistics & Epidemiology, School of Public Health and Health Sciences, University of Massachusetts Amherst, Amherst, Massachusetts, United States

**Source of Funding:** None

**Conflict of interest:** None

**Acknowledgements:** The authors acknowledge Drs. Nicholas G. Reich and Zhichao Jiang for their expert advice.

**Data availability:** Computing code to reproduce the study results is available in the Supplementary Materials.

**TITLE:** Evaluating the Impact of State-Level Public Masking Mandates on New COVID-19 Cases and Deaths in the United States: A Demonstration of the Causal Roadmap

**ABSTRACT:**

At a national-level, we sought to investigate the effect of public masking mandates on COVID-19 in Fall 2020. Specifically, we aimed to evaluate how the relative growth of COVID-19 cases and deaths would have differed if all states had issued a mandate to mask in public by September 1, 2020 versus if all states had delayed issuing such a mandate. To do so, we applied the Causal Roadmap, a formal framework for causal and statistical inference. The outcome was defined as the state-specific relative increase in cumulative cases and in cumulative deaths {21, 30, 45, 60}-days after September 1. Despite the natural experiment in state-level masking policies, the causal effect of interest was not identifiable. Nonetheless, we specified the target statistical parameter as the adjusted rate ratio (aRR): the expected outcome with early implementation divided by the expected outcome with delayed implementation, after adjusting for state-level confounders. To minimize strong estimation assumptions, primary analyses used targeted maximum likelihood estimation (TMLE) with Super Learner. After 60-days and at a national-level, early implementation was associated 9% reduction in new COVID-19 cases (aRR: 0.91; 95%CI: 0.88-0.95) and a 16% reduction in new COVID-19 deaths (aRR: 0.84; 95%CI: 0.76-0.93). Although lack of identifiability prohibited causal interpretations, application of the Causal Roadmap facilitated estimation and inference of statistical associations, providing timely answers to pressing questions in the COVID-19 response.

Despite demonstrated biological efficacy and official recommendations from the Centers of Disease Control and Prevention (CDC) (1–4), there has been wide variability in state-level policies for mask wearing while in public to reduce COVID-19 transmission (5,6). A survey conducted by the Delphi Group suggested a strong state-level correlation between mask wearing and COVID-19 case rates (7). While these correlation-based results are encouraging, failing to appropriately control for common causes of the policy and outcomes (i.e., confounders) can yield to misleading results (e.g., (8)).

The majority of studies aiming to evaluate the impact of masking mandates on COVID-19 in the United States (US) have relied on epidemic modeling (9). Fewer studies have employed "causal inference" techniques (10–13). For example, Lyu and Wehbly employed an event study design, which is similar to difference-in-differences, to investigate the impact of state-level masking mandates on confirmed COVID-19 cases from March through May 2020 (10). They found that mandates for public masking were strongly associated with reductions in county-level COVID-19 growth rates, but mandates for employee-only masking were not. In contrast, using an econometric approach based on linear structural equations, Chernozhukov et al. found that early implementation of a national mandate for employee-only masking could have saved upwards of 47,000 US lives by the end of May 2020 (11).

In this manuscript, we applied the Causal Roadmap (14–24) with the goal of evaluating the effect of delays in state-level public masking mandates on the relative growth of COVID-19 cases and deaths in the 50 US states from September 1 to October 31, 2020. The steps of the Causal Roadmap are 1) defining the research question, 2) specifying the causal model, 3) specifying the causal parameter using counterfactuals, 4) linking the observed data to causal model, 5) identifying the statistical estimand, 6) statistical estimation and inference, and 7) interpretation and discussion. Its application here demonstrates many of the challenges and potential solutions, commonly arising in studies of COVID-19 policies and interventions.

**THE RESEARCH QUESTION (Roadmap Step 1)**

We aimed to evaluate the effect of state-level public masking mandates on subsequent COVID-19 cases and deaths in the 50 US states. Throughout, we focused on the strongest masking policy, assuming that it would lead to the greatest impact. Specifically, we defined the exposure $A$ as an indicator that a state issued a mandate requiring masks or cloth face coverings in public indoor and outdoor spaces when it was not possible to maintain at least 6-feet distance by the target date (described next). In other words, $A=1$ for a state that had this masking policy in place by the target date (hereafter called "early"), and $A=0$ for a state that implemented this masking policy after the target date or never implemented the policy (hereafter called "delayed"). There were no states who issued and subsequently lifted the masking mandate before the target date. In other words, "early" acting states both issued and maintained the mandate.

In the primary analysis, the target date was September 1, 2020, selected to correspond roughly with the start of the school year for K-12 schools and higher education. Given the anticipated shift in behavior from summertime to classroom-based activities for many US residents and their families, we hypothesized that having the masking mandate in place before this target date would help limit subsequent spread of COVID-19. We also conducted a secondary analysis with the exposure defined as having issued the masking mandate before stay-at-home (i.e., shelter-in-place) orders were terminated. The target date for states that never issued a stay-at-home order was set to the median termination date among states that did: May, 15, 2020. In this secondary analysis, the target date varied by state.

To minimize the impact of measured and unmeasured influences of the COVID-19 trajectory between states, we focused our outcome definition on relative changes, allowing each state to be its own reference. Specifically, the outcome $Y$ was the state-specific COVID-19 relative growth, defined as the cumulative number of confirmed cases (deaths) a set number of days after the target date, divided by the cumulative number of confirmed cases (deaths) on the target date. Therefore, the outcome $Y$ was always greater than or equal to 1. To examine shorter- and longer-term associations, we considered outcome periods at {21, 30, 45, and 60}-days after the target date.  Altogether, we aimed to evaluate the state-level effect of having the public masking mandate in place prior to September 1, 2020 on subsequent COVID-19 cases and deaths through October 31, 2020. We hypothesized that the relative reduction in the COVID-19 outcomes associated with early implementation would grow over time.

**THE CAUSAL MODEL (Roadmap Step 2)**

To address common causes of the exposure and outcome, we considered state-level variables that a policy-maker might consult when deciding whether to enact and maintain the masking mandate and that might also influence subsequent COVID-19 outcomes. These measured confounders, denoted $W$, included population demographics (e.g., distributions of age and ethnicity), socio-economic measures, prevalence of co-morbidities, measures of population density, commuting patterns, and political leaning (indicator that the Republican party won the majority vote in the 2016 Presidential election). Additionally, prior public health policies (e.g., gathering restrictions, stay-at-home orders) and recently observed COVID-19 outcomes per-capita (e.g., COVID-19 tests, cases, and deaths) were likely to impact both the state government's urgency as well as their residents' behavior. Finally, as a proxy for prior response to COVID-19 policies, we included changes in Google's residential mobility indicators 7- and 14-days before the target date (25). A full list of potential confounders is provided in eAppendix 1.

Causal models are useful tools for representing the relationships between key variables in our study. As detailed below, causal models also facilitate the derivation of counterfactual outcomes corresponding to our research question, the assessment of identifiability, and the definition of a statistical model avoiding unsubstantiated assumptions (14,15,24,26). Therefore, we specified the following structural causal model to present the data generating process for each state, including its measured confounders $W$, exposure $A$, and outcome $Y$ (14,26):

$$W = f_W(U_W)$$

$$A = f_A(W, U_A)$$

$$Y = f_Y(W, A, U_Y)$$

where $(f_W, f_A, f_Y)$ were the non-parametric structural equations and $(U_W, U_A, U_Y)$ were the unmeasured factors contributing the confounders, exposure, and outcome, respectively. The corresponding causal graph is given in **Figure 1**. These causal models were specified at the state-level and encoded the assumption of no interference between states (i.e., the outcome for one state was not impacted by another's policy).

Importantly, we did not specify the functional form of the structural equations. Equally importantly, we did not specify any independence assumptions between the unmeasured factors; thus, we encoded that there were unmeasured common causes of the confounders $W$, the exposure $A$, and the outcome $Y$. (On the causal graph in **Figure 1**, unmeasured common causes were indicated by the dashed double-headed arrows.) A key unmeasured confounder was the state's COVID-19 landscape; specifically, the epidemic trajectory near the target date (e.g., upslope, apex, downslope, low plateau) could impact whether the masking policy was in place as well as the relative growth of subsequent COVID-19 outcomes. While including recently observed COVID-19 outcomes in the confounders $W$ partially accounted for the past epidemic trajectory, the anticipated future trajectory likely influenced state policies and residents' behavior. Examples of other unmeasured confounders included perceived or actual compliance with previous public health policies and the strength of the state's public health department.

**CAUSAL PARAMETER (Roadmap Step 3):**

Recall our goal of evaluating the effect of early versus delayed public masking mandates on subsequent COVID-19 cases and deaths at a national-level. To translate this goal into a well-specified causal parameter, we defined $Y_a$ as the counterfactual outcome for a state if the exposure-level were $A = a$. For example, $Y_1$ was the relative increase in the COVID-19 case rate for a state if possibly contrary-to-fact they implemented the masking mandate by the target date, while $Y_0$ was the relative increase in the COVID-19 case rate for a state if possibly contrary-to-fact they failed to implement the masking mandate by the target date. These counterfactual outcomes were derived by deterministically setting the exposure $A$ equal to 1 and to 0 in the above structural causal model (14,26). Counterfactual outcomes for COVID-19 deaths were defined and derived analogously.

Using these counterfactual outcomes, we then specified our target causal parameter as the causal rate ratio $CRR = \mathbb{E}[Y_1] \div \mathbb{E}[Y_0]$ in the primary approach and considered the causal rate difference $CRD = \mathbb{E}[Y_1] - \mathbb{E}[Y_0]$ in secondary analyses. For both parameters, the expectations are over the distribution of state-level, counterfactual outcomes. In words, these causal parameters correspond to the ratio and difference in the expected relative growth in COVID-19 cases (deaths) if all 50 states had early versus delayed implementation of the masking policy. In reality, of course, we did not observe both counterfactual outcomes, since a state either did or did not implement the masking mandate by the target date.

**OBSERVED DATA & STATISTICAL MODEL (Roadmap Step 4):**

We obtained state-level confounders $W$ from the Bureau of Transportation Statistics (27), Iowa Community Indicators Program (28), Kaiser Family Foundation (5), MIT Election Data and Science Lab (29), Google Mobility Report (25), and the US Census Bureau (30). Data on COVID-19 policies were collected from the GitHub repository maintained by the COVID-19 State Policy Team at the University of Washington, with verification as needed from the Kaiser Family Foundation and state government websites (5,6). For each policy, the issued, enacted, expired, and end dates were collected. From these, we created binary indicators to represent whether a policy was in place by a certain date. Given variability in their strictness, masking mandates were categorized into 3 levels: limited to specific public settings, more broadly required indoors and in enclosed spaces, and generally required for both indoor and outdoor public spaces where 6-feet distance cannot be maintained. However, as previously discussed, only the strictest public masking mandate, level-3, was considered in this study. Finally, time series data on COVID-19 tests, cases, and deaths were collected from the COVID Tracking Project (31).

The observed data $O$ were at the state-level and consisted of the measured confounders $W$, the masking mandate indicator $A$, and the relative growth outcome $Y$. We assumed the observed data were generated according to a process compatible with the causal model, which did not place any restrictions on the set of possible distributions of the observed data. Thereby, the statistical model was non-parametric (14,24).

**IDENTIFICATION (Roadmap Step 5):**

Next, we assessed the assumptions needed to express the causal parameter as some function of the observed data distribution. In previous steps of the Roadmap, we already made the temporality, no interference, and consistency assumptions (14). For identifiability, we would additionally need there to be no unmeasured common causes of the exposure $A$ and the outcome $Y$. However, this assumption did not hold in our causal model (Roadmap Step 2); therefore, due to unmeasured confounding, we concluded that the causal effect of interest was not identifiable despite the natural experiment occurring between states. Nonetheless, through the G-computation formula (32), we specified our statistical estimand as the adjusted rate ratio:

$$aRR = \frac{\mathbb{E}[\mathbb{E}(Y|A=1,W)]}{\mathbb{E}[\mathbb{E}(Y|A=0,W)]}$$

In words, $\mathbb{E}[\mathbb{E}(Y|A = a, W)]$ was the expected outcome, given mandate implementation *A=a* and the measured confounders *W*, and then standardized with respect to the confounder distribution in the population. A ratio less than 1 represented that early implementation of the masking mandate was associated with reductions in relative growth of COVID-19 at the national-level. In secondary analyses, we examined the adjusted rate difference: $aRD = \mathbb{E}[\mathbb{E}(Y|A = 1, W)] - \mathbb{E}[\mathbb{E}(Y|A = 0, W)]$.

For either statistical estimand to be well-defined, we needed an additional condition on data support, sometimes called the "positivity assumption", "overlap", and the "experimental treatment assignment assumption": $min_{a \in \{0,1\}} \mathbb{P}(A = a|W = w) > 0$ for all possible *w* (33). In this application, the target population was the same as the sample; in other words, there are *N*=50 states for which we wanted to make inferences. Therefore, in this study, "structural" positivity violations occurring if early or delayed implementation was theoretically impossible for some state-level confounder values were equivalent to "chance" positivity violations occurring due to finite sample sizes (33). Additionally, given the high dimension of *W,* we had to balance confounding control with bias arising from positivity violations. To do so, we reduced the adjustment set *W* through screening based on univariate associations with the outcome, an approach which should help screen-out instrumental variables and reduce data sparsity, but also risks exclusion of a confounder weakly associated with the outcome.

**STATISTICAL ESTIMATION AND INFERENCE (Roadmap Step 6)**

In line with recommendations for studying COVID-19 (34) and desirable asymptotic properties that often translate into improved finite sample performance (e.g., (35–40)), we used targeted maximum likelihood estimation (TMLE) with Super Learner for statistical estimation and inference (14). TMLE requires estimation of both the expected outcome, given the exposure and the confounders $\mathbb{E}(Y|A = a, W)$, and the conditional probability of the exposure, given the confounders $\mathbb{P}(A = 1|W)$. If either is estimated consistently, TMLE will yield a consistent estimate and is double robust. If both are consistently estimated at reasonable rates, TMLE is efficient with lowest asymptotic variance among a large class of estimators. Importantly, TMLE allows for the utilization of machine learning to avoid strong parametric assumptions, while obtaining valid statistical inference (e.g., 95% confidence intervals with nominal coverage) under fairly mild conditions (14,41). Practical descriptions of the algorithm are available in (42,43), among others.

Within TMLE, we used the Super Learner, an ensemble method, to avoid strong estimation assumptions and to respect the non-parametric statistical model (44). Using 10-fold cross-validation, Super Learner built the best weighted combination of predictions from the following algorithms: the empirical mean, generalized additive models (*gam*), recursive partitioning and regression trees (*rpart*), extreme gradient boosting (*xgboost*), and multivariate adaptive regression splines (*earth*), each paired with a screening algorithm based on univariate correlations with the outcome (45–49). For demonstration we also implemented the unadjusted estimator, as the contrast in the average outcome among early states and the average outcome among delayed states. We obtained statistical inference via the estimated influence curve. All analyses were conducted in *R* v4.0.2 with the *ltmle* package (50,51). Full computing code is available in eAppendix 2.

**INTERPRETATION AND DISCUSSION OF RESULTS (Roadmap Step 7)**

Twenty-five states had the state-level public masking mandate in place by September 1, 2020, while 25 did not. Among the latter, 7 states never implemented a masking mandate; 12 states had a less strict masking mandate, and 6 states implemented the desired mandate after the target date. There were some notable differences between groups (**Table 1**). In the early acting states, there was a higher percentage of Black and Hispanic residents, higher population density, fewer Republican voters, and more per-capita COVID-19 tests, cases, and deaths leading up to September 1, 2020. States with early implementation were also more likely to have previously issued orders to stay-at-home and for school masking.

Within three weeks of the target date, the expected relative growth of COVID-19 outcomes with and without early implementation of the masking mandate started to diverge (**Table 2; eFigure 1**). Specifically, the expected change in confirmed cases after 21 days was 1.16 (95%CI: 1.14-1.19) with early implementation and 1.20 (95%CI: 1.17-1.24) with delayed implementation for an adjusted ratio of 0.96 (95%CI: 0.95-0.98). After one month, the adjusted ratio was 0.95 (95%CI: 0.92-0.97) and continued to decline over time (**Figure 2A**). After 45 and 60 days, the ratios were 0.92 (95%CI: 0.90-0.95) and 0.91 (95%CI: 0.88-0.95), respectively. Thus, early implementation of the state-level public masking mandate was associated with a 9% relative reduction in the relative growth of COVID-19 cases after two-months.

Similar patterns but larger long-term associations were observed for COVID-19 deaths (**Figure 2B**). Specifically, the expected change in COVID-19 deaths after 60 days was 1.44 (95%CI: 1.29-1.58) with early implementation and 1.71 (95%CI: 1.51-1.90) with delayed implementation for an adjusted ratio of 0.84 (95%CI: 0.76-0.93). Thus, early implementation of the state-level public masking mandate was associated with a 16% relative reduction in the change of COVID-19 deaths after two months. For both endpoints, similar patterns were seen on the absolute scale (**Table 2**).

As expected, however, associations were considerably larger when using an unadjusted approach (**eTable 1**). For example, the unadjusted ratios at 60-days were 0.72 (95% CI: 0.61-0.85) for cases and 0.71 (95% CI: 0.59-0.85) for deaths. Thus, failing to adjust for state-level measured confounders increased the estimated two-month associations from 9% to 28% for the case outcome and from 16% to 29% for the death outcome.

In the secondary analysis defining the target date according to the relaxation of stay-at-home orders, there were only 8 states considered "early" implementers: Connecticut, Delaware, Illinois,

Massachusetts, Maine, New Mexico, New York, and Rhode Island. This secondary analysis resulted in qualitatively similar conclusions to the primary one: early implementation of the public masking mandate was associated with reduced growth in COVID-19 and the associations strengthened over time (**eFigure 2**). However, the longer-term associations were much stronger for confirmed cases and weaker for deaths. Specifically, the 60-day adjusted ratios were 0.71 (95%CI: 0.66-0.77) for the case endpoint and 0.92 (95%CI: 0.87-0.98) for the death endpoint. These results highlighted how asking different research questions (i.e., mandate by September 1, 2020 vs. by lifting stay-at-home) can yield qualitatively similar, but quantitatively different conclusions. Altogether, implementation of a state-level public masking mandate prior to lifting stay-at-home was associated with a 29% relative reduction in new cases and a 8% relative reduction in new deaths after two months. As before, larger associations were seen when using an unadjusted estimation approach (**eTable 2**).

There were several limitations to our findings. First, our analyses may be subject to bias due to unmeasured confounding, incomplete control for measured confounders, and complex dependence. As previously discussed, examples of unmeasured common causes of the masking policy and COVID-19 outcomes included the (perceived) epidemic trajectory, the (perceived) compliance with prior public health policies, and the strength of the state's public health department. There are likely many other factors; indeed, the spatial-temporal waxing-and-waning of COVID-19 is still poorly understood. Nonetheless, given the apparent differences between early and delayed states on measured confounders (e.g., population demographics and prior COVID-19 outcomes as shown in **Table 1**), we expected that our adjusted results were closer to the true effect than the unadjusted results. Of course, it is always possible that the unmeasured confounders balanced out the impact of the measured confounders.

Secondly, our adjusted analyses may be subject to bias due to incomplete control for measured confounding. As commonly occurring, the set of potential confounders was high-dimensional, especially relative to our sample size (N=50 states). To improve data support and reduce the potential for bias due to violations of the positivity assumption (33), we reduced the confounder set based on univariate associations with the outcome. This approach should help avoid controlling for instrumental variables but also risks exclusion of a key confounder that is weakly associated with the outcome. In this application, the minimum and maximum of the estimated propensity scores were far from 0 and 1 (**eTable 3**).

Third, our causal model explicitly assumed independence between states. Due to the infectious nature of COVID-19 and travel between states, this assumption was likely to be violated. In cluster randomized trials, such interference is expected to bias the effect towards the null (52); however, in settings with complex network dependence, the impact of interference can be unpredictable (53). In all scenarios, ignoring interference and other sources of dependence will cause our confidence intervals to be overly precise. Future work could partially relax this assumption by adjusting for the pre-exposure COVID-19 outcomes of neighboring states.

Altogether, this work could be strengthened by the application of quantitative bias analyses (54) and other sensitivity analyses to formally gauge the divergence between our wished-for causal parameters (e.g., $\mathbb{E}(Y_a)$) and the corresponding statistical parameters (e.g, $\mathbb{E}[\mathbb{E}(Y|a, W)]$). Indeed, there were many concerns about the causal assumptions needed for identifiability, including unmeasured confounding, positivity violations, and interference. However, by following the Causal Roadmap, we avoided being paralyzed into inaction. Instead, we specified statistical parameters that would equal causal parameters if the identifiability assumptions did, in fact, hold. Furthermore, we estimated these statistical quantities using state-of-the art methods to minimize statistical bias (i.e., the divergence between the statistical target and its estimate). Finally, application of the Causal Roadmap helped ensure that we interpreted the resulting point estimates and inferences appropriately – specifically as associations providing best answers to critical policy decisions.

Additional challenges of this application include imperfectly defined COVID-19 endpoints and the examination of a single masking mandate. Our primary analyses focused on COVID-19 cases, which may be subject to measurement error due to the availability of testing and the delay between sample collection and result reporting. During the time period of primary interest (September 1 to October 31, 2020) testing was widely available and reporting systems were well-established. Nonetheless, it is important to interpret these results in terms of confirmed cases, which given the high prevalence of asymptomatic infections were likely to only capture a fraction of all COVID-19 infections. Additionally, variability in a state's case ascertainment rate could introduce measurement error into the outcome. Importantly, however, similar results were observed for the secondary analysis investigating associations with COVID-19 deaths, an endpoint expected, but not guaranteed, to be more robust to these measurement issues. Finally, we focused on implementation of the strictest public masking mandate. It would be interesting to investigate if and how the association varied by type of masking mandate.

Evaluating the national impact of state-level masking mandates by the approximate start of the school year is just one of the possible research questions of interest. Indeed, our secondary analysis examined associations with respect to the lifting of stay-at-home orders. Furthermore, the framework that we presented can be applied to study the effects of longitudinal exposures (e.g., the impact of daily public health decisions), dynamic exposures (e.g., implementation based on state-specific characteristics), stochastic exposures (e.g., shifts in the distribution of delays to implementation), and much more! Indeed, a recent Google search returned over 400,000 abstracts on COVID-19. We hope that our tutorial on the Causal Roadmap will improve the design and analysis of these studies, especially with respect to the transparent statement and careful evaluation of identifiability assumptions.

**CONCLUSION**

In this study, our objective was to demonstrate the application of the Causal Roadmap (15,17) to evaluate the state-level effect of masking mandates on COVID-19 outcomes in Fall 2020 in the United States. We specified a causal model to represent our background knowledge, intervened on that causal model to derive counterfactual outcomes, formally addressed identifiability (or lack thereof), and estimated the association of having a state-level, public masking mandate in place prior to September 1, 2020, the approximate start of the school year, on the relative growth of COVID-19 cases and deaths. We evaluated outcomes at four different periods (i.e., 21, 30, 45, 60-days after the target date) and observed that associations increased meaningfully over time.

Our study builds on growing evidence that prior to widespread vaccination, public masking mandates are associated with population-level reductions in COVID-19 spread. Our results are in line with those of Lyu and Wehby who found the association between state-level public masking mandates and COVID-19 growth rates increased over time (10). Our results are also in line with a regression-based study by Krishnamachari et al., who found that longer delays between the CDC guidance and state-level masking mandates were associated with higher cumulative case rates (55).

To the best of our knowledge, this is the first work employing the Causal Roadmap and TMLE to evaluate the impact of state-level policies for masking on COVID-19 cases and deaths at a national-level. Although the causal effect of interest was not identifiable, we still specified a statistical estimand that best answered our research question. Furthermore, we estimated this parameter using state-of-the-art methods to avoid strong parametric assumptions. Altogether, we believe a particular strength of the Causal Roadmap is elaborating and critically evaluating the assumptions needed for causal inference and statistical inference.

**Table 1:** Summary of baseline characteristics for the 50 US states, overall and by exposure group: "early" if the public masking mandate was in place by September 1, 2020 and "delayed" otherwise. All metrics are given in median (25% quantile, 75% quantile) unless noted. The number of states that had Republican majority votes, issued Stay-at-Home, Gathering Restrictions, or School Masking are given in counts (percentage).

| | All (N=50) | Early Masking (N=25) | Delayed Masking (N=25) |
|---|---|---|---|
| **Population Demographics** | | | |
| Black or African American (%) | 7 (3.2, 14.2) | 9.9 (5.7, 14) | 4.2 (2, 15.3 |
| Hispanic (%) | 9.4 (5.1, 13.8) | 11.8 (5.1, 17.1) | 7 (4.3, 10.6) |
| Mixed Race (%) | 2.2 (1.9, 2.5) | 2.1 (1.9, 2.5) | 2.2 (2, 2.5) |
| Caucasian (%) | 71.8 (59.2, 79.7) | 68.5 (55.6, 75.9) | 78.3 (63.1, 82) |
| Median Age | 38.4 (37.1, 39.5) | 38.8 (37.7, 39.9) | 38.2 (36.7, 39.1) |
| Smoker (%) | 17.1 (15, 19.3) | 17 (14.1, 19.3) | 17.2 (15.6, 19.3) |
| **Political Leaning** | | | |
| Republican | 30 (60%) | 10 (40%) | 20 (80%) |
| **Population Density & Urbanicity** | | | |
| Total Population | 4551910 (1847981, 7207423) | 4663616 (2922849, 9957488) | 3918137 (1422029, 6090062) |
| Population Density (people per km$^2$) | 41.2 (17.7, 84.7) | 67.9 (24.6, 160.7) | 26.8 (9.6, 62.3) |
| Urbanicity in 2010 (%) | 73.8 (65.1, 87) | 81 (73.2, 88) | 66.4 (64, 75.1) |
| Public Transportation Usage (%) | 1.4 (0.8, 3.4) | 1.8 (0.9, 5.8) | 1.2 (0.8, 2) |
| **Prior COVID-19 Outcomes** (per 100,000 residents) | | | |
| Confirmed Cases 30 days prior | 1161.9 (854.1, 1629.8) | 1390.8 (888, 1717.7) | 1036.1 (851.2, 1452.3) |
| Confirmed Cases 14 days prior | 1359 (972.7, 1921.1) | 1616.1 (981.8, 1958.3) | 1258.9 (969.6, 1690.4) |
| Confirmed Cases 7 days prior | 1417.8 (1010.8, 1996.6) | 1719.3 (1016.6, 2022.7) | 1362.5 (1008.8, 1829) |
| Deaths 30 days prior | 27.2 (14.7, 51.9) | 44.8 (21.8, 64.8) | 17.4 (11.7, 31) |
| Deaths 14 days prior | 31.6 (17.6, 56) | 47.7 (24.5, 71.2) | 21.4 (16.2, 32) |
| Deaths 7 days prior | 32.5 (18.9, 57.6) | 48.8 (25.6, 77.5) | 23.6 (18.6, 33.6) |
| **Prior COVID-19 Policies** | | | |
| Implemented Stay-at-Home | 43 (86%) | 24 (96%) | 19 (76%) |
| Implemented Gathering Restrictions | 49 (98%) | 25 (100%) | 24 (96%) |
| Implemented School Masking | 28 (56%) | 18 (72%) | 10 (40%) |
| **Changes in Mobility** | | | |
| Mobility Change 14 days prior (%) | 8.5 (7, 10) | 9 (7, 11) | 7 (6, 10) |
| Mobility Change 7 days prior (%) | 9 (7, 11) | 9 (8, 11) | 8 (6, 10) |

**Table 2:** <u>For the target date of September 1, 2020,</u> point estimates (95% confidence intervals (CI)) for the expected outcome under early implementation $\mathbb{E}[\mathbb{E}(Y|A = 1, W)]$, the expected outcome under delayed implementation $\mathbb{E}[\mathbb{E}(Y|A = 0, W)]$, their ratio (RR), and their difference (RD) estimated using TMLE with Super Learner.

|  | At | Early (95% CI) | Delayed (95% CI) | RR (95% CI) | RD (95% CI) |
|---|---|---|---|---|---|
| **Cases** | 21 days | 1.16 (1.14, 1.19) | 1.20 (1.17, 1.24) | 0.96 (0.95, 0.98) | -0.04 (-0.06, -0.03) |
|  | 30 days | 1.25 (1.20, 1.29) | 1.32 (1.26, 1.37) | 0.95 (0.92, 0.97) | -0.07 (-0.10, -0.04) |
|  | 45 days | 1.44 (1.35, 1.52) | 1.56 (1.45, 1.67) | 0.92 (0.90, 0.95) | -0.12 (-0.16, -0.08) |
|  | 60 days | 1.76 (1.61, 1.90) | 1.92 (1.73, 2.12) | 0.91 (0.88, 0.95) | -0.16 (-0.24, -0.09) |
| **Deaths** | 21 days | 1.13 (1.09, 1.16) | 1.19 (1.14, 1.23) | 0.95 (0.92, 0.98) | -0.06 (-0.10, -0.02) |
|  | 30 days | 1.21 (1.15, 1.27) | 1.26 (1.20, 1.33) | 0.96 (0.91, 1.00) | -0.05 (-0.11, 0.00) |
|  | 45 days | 1.29 (1.21, 1.36) | 1.46 (1.36, 1.57) | 0.88 (0.82, 0.94) | -0.18 (-0.27, -0.08) |
|  | 60 days | 1.44 (1.29, 1.58) | 1.71 (1.51, 1.90) | 0.84 (0.76, 0.93) | -0.27 (-0.43, -0.11) |

**Figure 1:** Causal graph representing the data generating process for the study, with *W* as measured confounders, *A* as the masking policy, *Y* as the COVID-19 outcome, and *(U_W, U_A, U_Y)* as the unmeasured background variables for *(W,A,Y)*, respectively. Directed arrows represent direct causes, and dashed double-headed arrows represent dependence between the background factors.

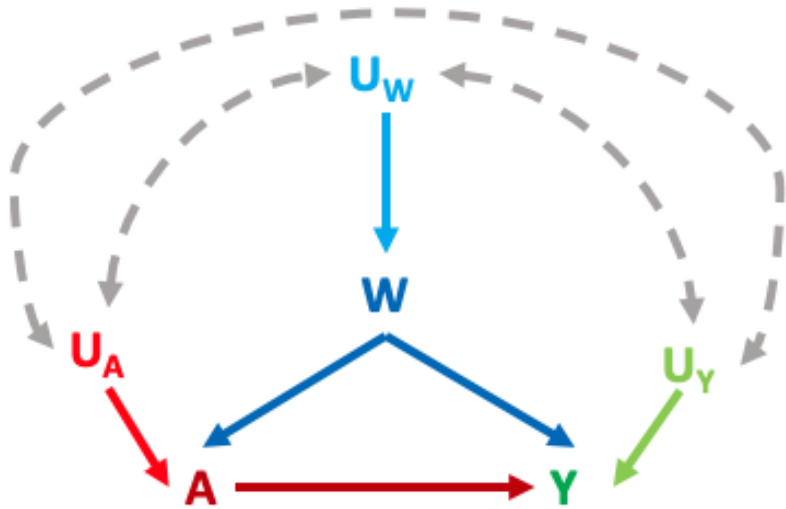

**Figure 2:** For the target date of September 1, 2020, point estimates (95% confidence intervals) for the ratios in the expected relative growth of COVID-19 cases (A) and deaths (B) under early and delayed implementation of the public masking mandate over time.

(A)

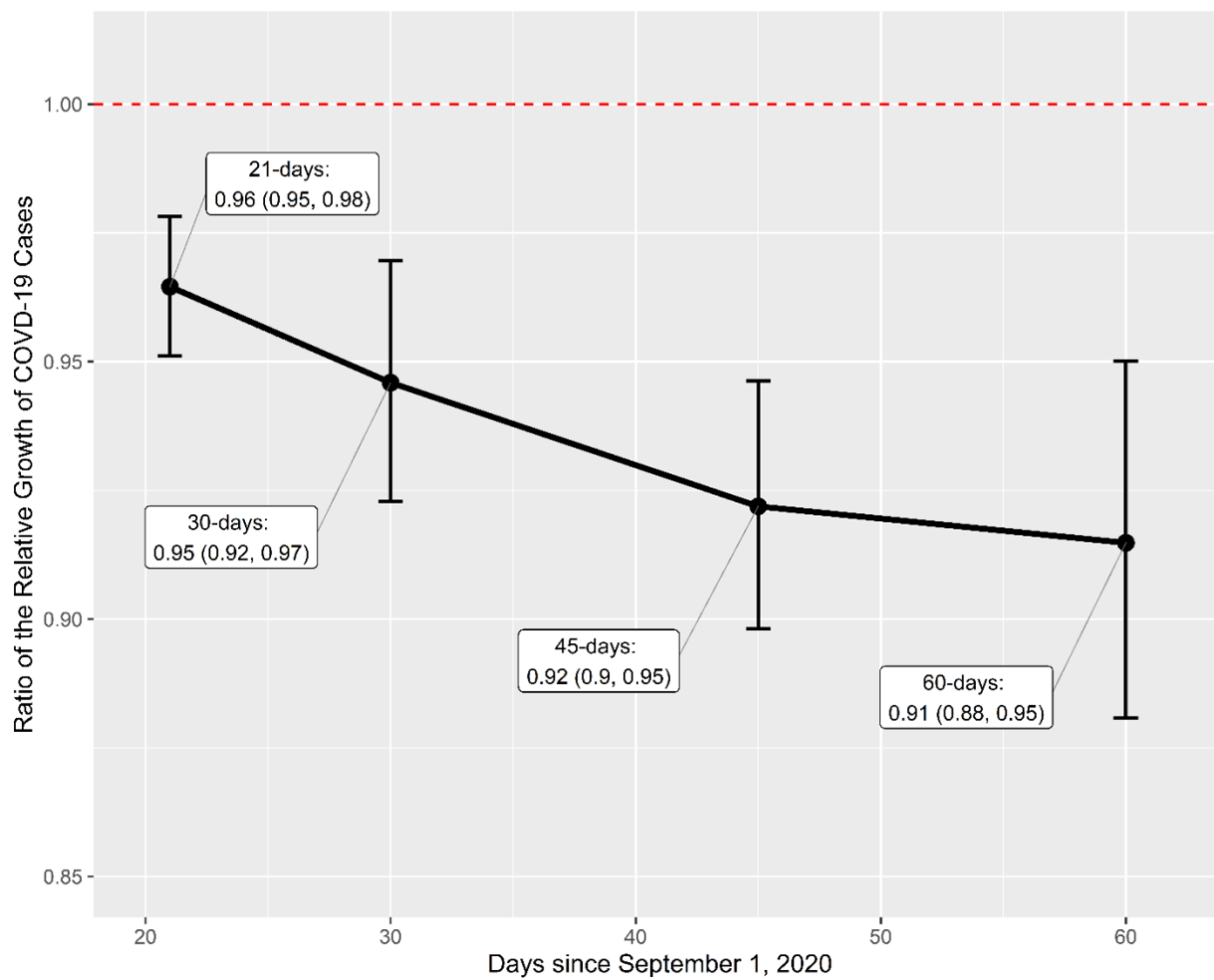

(B)

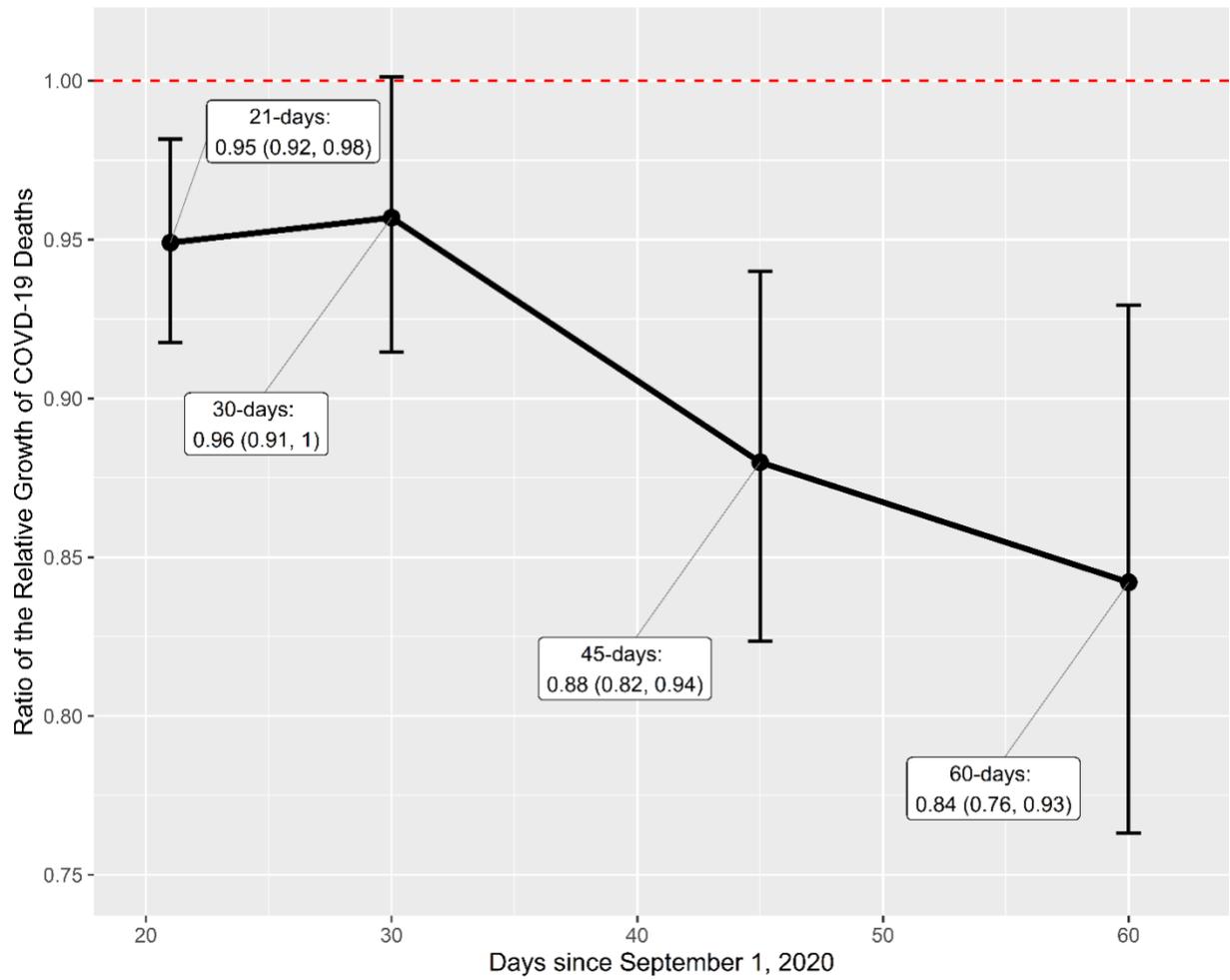

***SUPPLEMENTARY MATERIALS: Evaluating the Impact of State-Level Public Masking Mandates on New COVID-19 Cases and Deaths in the United States: A Demonstration of the Causal Roadmap***

**eAppendix 1:** Full list of state-level confounders considered. Recall in the primary approach, Super Learner was implemented after reducing the potential adjustment set based on univariate correlations with the timepoint-specific outcome.

(1) Percent of population that is over 65 years old,
(2) Percent of population that is Black or African American,
(3) Percent of population that is Hispanic,
(4) Percent of population that is Asian,
(5) Percent of population that is Mixed Race,
(6) Percent of population that is Caucasian,
(7) Median Age,
(8) Percent of Household with income below poverty level,
(9) Percent of People with Income below poverty level,
(10) Percent of Smoker,
(11) Percent of Diabetic people,
(12) Population Density (people per $km^2$),
(13) Percent of people that drive to work,
(14) Percent of people that work from home,
(15) Percent of people that take public transportation to work,
(16) Percent of people that bike to work,
(17) Percent of people that walk to work,
(18) Percent of people with other means of commute to work,
(19) Total population,
(20) Republican,
(21) Cumulative Confirmed Positive Cases per 100,000-residents at 30 days before target date,
(22) Cumulative Confirmed Positive Cases per 100,000-residents at 14 days before target date,
(23) Cumulative Confirmed Positive Cases per 100,000-residents at 7 days before target date,
(24) Cumulative Deaths per 100,000-residents at 30 days before target date,
(25) Cumulative Deaths per 100,000-residents at 14 days before target date,
(26) Cumulative Deaths per 100,000-residents at 7 days before target date,
(27) Cumulative Total COVID-19 Tests per 100,000-residents at 30 days before target date,
(28) Cumulative Total COVID-19 Tests per 100,000-residents at 14 days before target date,
(29) Cumulative Total COVID-19 Tests per 100,000-resident sat 7 days before target date,
(30) Ever implemented a Shelter-in-place/Stay-at-Home policy,
(31) Ever implemented a Gathering Restriction policy,
(32) Ever implemented a Restaurant Restriction policy,
(33) Ever implemented a Non-Essential Business Closure policy,
(34) Ever implemented another Business Closure policy,
(35) Ever implemented a Business Masking policy,
(36) Ever implemented a School Masking policy,
(37) Percent change in Residential Area Mobility from baseline, at 14 days before target date,
(38) Percent change in Residential Area Mobility from baseline, at 7 days before target date

**eAppendix 2: Computing code**

R code for the primary analysis is available at:

https://github.com/angus-wong/Evaluating-the-Impact-of-State-Level-Public-Masking-Mandates-on-New-COVID-19-Cases-and-Deaths

**eTable 1a:** For the target date of September 1, 2020 and confirmed COVID-19 cases, point estimates (95% confidence intervals) for the expected outcome under early implementation $\mathbb{E}[\mathbb{E}(Y|A=1,W)]$, the expected outcome under delayed implementation $\mathbb{E}[\mathbb{E}(Y|A=0,W)]$, their ratio (RR), and their difference (RD) using TMLE and the unadjusted estimator. (In the unadjusted estimator, $W$ is the empty set *{}*).

|          | At      | Early (95% CI)    | Delayed (95% CI)  | RR (95% CI)       | RD (95% CI)          |
|----------|---------|-------------------|-------------------|-------------------|----------------------|
| TMLE     | 21 days | 1.16 (1.14, 1.19) | 1.2 (1.17, 1.24)  | 0.96 (0.95, 0.98) | -0.04 (-0.06, -0.03) |
|          | 30 days | 1.25 (1.2, 1.29)  | 1.32 (1.26, 1.37) | 0.95 (0.92, 0.97) | -0.07 (-0.1, -0.04)  |
|          | 45 days | 1.44 (1.35, 1.52) | 1.56 (1.45, 1.67) | 0.92 (0.9, 0.95)  | -0.12 (-0.16, -0.08) |
|          | 60 days | 1.76 (1.61, 1.9)  | 1.92 (1.73, 2.12) | 0.91 (0.88, 0.95) | -0.16 (-0.24, -0.09) |
| Unadjusted | 21 days | 1.13 (1.1, 1.16) | 1.25 (1.2, 1.29)  | 0.91 (0.86, 0.95) | -0.12 (-0.17, -0.06) |
|          | 30 days | 1.19 (1.15, 1.23) | 1.39 (1.3, 1.47)  | 0.86 (0.8, 0.92)  | -0.2 (-0.29, -0.1)   |
|          | 45 days | 1.33 (1.27, 1.4)  | 1.7 (1.52, 1.87)  | 0.79 (0.7, 0.88)  | -0.36 (-0.55, -0.18) |
|          | 60 days | 1.54 (1.44, 1.65) | 2.16 (1.83, 2.49) | 0.72 (0.61, 0.85) | -0.61 (-0.96, -0.27) |

**eTable 1b:** For the target date of September 1, 2020 and COVID-19 deaths, point estimates (95% confidence intervals) for the expected outcomes under early implementation $\mathbb{E}[\mathbb{E}(Y|A=1,W)]$, the expected outcome under delayed implementation $\mathbb{E}[\mathbb{E}(Y|A=0,W)]$, their ratio (RR), and their difference (RD) using TMLE and the unadjusted estimator. (In the unadjusted estimator, $W$ is the empty set *{}*).

|          | At      | Early (95% CI)    | Delayed (95% CI)  | RR (95% CI)       | RD (95% CI)          |
|----------|---------|-------------------|-------------------|-------------------|----------------------|
| TMLE     | 21 days | 1.13 (1.09, 1.16) | 1.19 (1.14, 1.23) | 0.95 (0.92, 0.98) | -0.06 (-0.1, -0.02)  |
|          | 30 days | 1.21 (1.15, 1.27) | 1.26 (1.2, 1.33)  | 0.96 (0.91, 1)    | -0.05 (-0.11, 0)     |
|          | 45 days | 1.29 (1.21, 1.36) | 1.46 (1.36, 1.57) | 0.88 (0.82, 0.94) | -0.18 (-0.27, -0.08) |
|          | 60 days | 1.44 (1.29, 1.58) | 1.71 (1.51, 1.9)  | 0.84 (0.76, 0.93) | -0.27 (-0.43, -0.11) |
| Unadjusted | 21 days | 1.1 (1.06, 1.15) | 1.21 (1.15, 1.27) | 0.91 (0.86, 0.98) | -0.1 (-0.18, -0.03)  |
|          | 30 days | 1.15 (1.08, 1.21) | 1.32 (1.23, 1.41) | 0.87 (0.8, 0.95)  | -0.17 (-0.28, -0.06) |
|          | 45 days | 1.22 (1.12, 1.32) | 1.53 (1.36, 1.7)  | 0.8 (0.7, 0.92)   | -0.31 (-0.5, -0.11)  |
|          | 60 days | 1.31 (1.17, 1.44) | 1.85 (1.58, 2.12) | 0.71 (0.59, 0.85) | -0.54 (-0.84, -0.24) |

**eTable 2a:** <u>For the target date corresponding to the state-specific relaxation of stay-at-home orders and confirmed COVID-19 cases,</u> point estimates (95% confidence intervals) for the expected outcome under early implementation $\mathbb{E}[\mathbb{E}(Y|A=1,W)]$, the expected outcome under delayed implementation $\mathbb{E}[\mathbb{E}(Y|A=0,W)]$, their ratio (RR), and their difference (RD) using TMLE and the unadjusted estimator. (In the unadjusted estimator, $W$ is the empty set $\{\}$).

| | At | Early (95% CI) | Delayed (95% CI) | RR (95% CI) | RD (95% CI) |
|---|---|---|---|---|---|
| TMLE | 21 days | 1.45 (1.38, 1.51) | 1.44 (1.36, 1.53) | 1 (0.98, 1.02) | 0 (-0.03, 0.03) |
| | 30 days | 1.51 (1.43, 1.58) | 1.68 (1.55, 1.81) | 0.9 (0.86, 0.94) | -0.17 (-0.25, -0.08) |
| | 45 days | 1.86 (1.72, 1.99) | 2.23 (1.96, 2.5) | 0.83 (0.77, 0.89) | -0.38 (-0.55, -0.2) |
| | 60 days | 2.21 (1.98, 2.44) | 3.11 (2.62, 3.6) | 0.71 (0.66, 0.77) | -0.9 (-1.2, -0.6) |
| Unadjusted | 21 days | 1.2 (1.09, 1.31) | 1.49 (1.4, 1.58) | 0.81 (0.72, 0.9) | -0.29 (-0.43, -0.15) |
| | 30 days | 1.27 (1.13, 1.42) | 1.74 (1.6, 1.88) | 0.73 (0.63, 0.84) | -0.47 (-0.67, -0.26) |
| | 45 days | 1.4 (1.18, 1.62) | 2.33 (2.04, 2.62) | 0.6 (0.49, 0.74) | -0.93 (-1.3, -0.56) |
| | 60 days | 1.56 (1.21, 1.9) | 3.33 (2.79, 3.87) | 0.47 (0.35, 0.62) | -1.78 (-2.42, -1.13) |

**eTable 2b:** <u>For the target date corresponding to the state-specific relaxation of stay-at-home orders and COVID-19 deaths,</u> point estimates (95% confidence intervals) for the expected death under early implementation $\mathbb{E}[\mathbb{E}(Y|A=1,W)]$, the expected outcome under delayed implementation $\mathbb{E}[\mathbb{E}(Y|A=0,W)]$, their ratio (RR), and their difference (RD) using TMLE and the unadjusted estimator. (In the unadjusted estimator, $W$ is the empty set $\{\}$).

| | At | Early (95% CI) | Delayed (95% CI) | RR (95% CI) | RD (95% CI) |
|---|---|---|---|---|---|
| TMLE | 21 days | 1.41 (1.33, 1.49) | 1.44 (1.34, 1.54) | 0.98 (0.93, 1.03) | -0.03 (-0.1, 0.05) |
| | 30 days | 1.54 (1.47, 1.62) | 1.6 (1.48, 1.73) | 0.96 (0.93, 1) | -0.06 (-0.12, 0) |
| | 45 days | 1.73 (1.61, 1.85) | 1.84 (1.66, 2.02) | 0.94 (0.9, 0.98) | -0.11 (-0.2, -0.03) |
| | 60 days | 1.93 (1.8, 2.06) | 2.09 (1.86, 2.32) | 0.92 (0.87, 0.98) | -0.16 (-0.29, -0.03) |
| Unadjusted | 21 days | 1.25 (1.09, 1.41) | 1.48 (1.37, 1.58) | 0.85 (0.73, 0.98) | -0.23 (-0.42, -0.03) |
| | 30 days | 1.32 (1.11, 1.54) | 1.65 (1.51, 1.79) | 0.8 (0.67, 0.97) | -0.32 (-0.58, -0.06) |
| | 45 days | 1.42 (1.14, 1.69) | 1.92 (1.72, 2.11) | 0.74 (0.59, 0.92) | -0.5 (-0.84, -0.16) |
| | 60 days | 1.5 (1.18, 1.82) | 2.2 (1.94, 2.46) | 0.68 (0.54, 0.87) | -0.69 (-1.1, -0.28) |

**eTable 3:** Summary of the estimated propensity scores for implementing the masking mandate by the target date of September 1, 2020.

| Outcome | At | Min. | 1st Qu. | Median | Mean | 3rd Qu. | Max. |
|---------|---------|-------|---------|--------|-------|---------|-------|
| Cases | 21 days | 0.340 | 0.376 | 0.464 | 0.500 | 0.649 | 0.728 |
|  | 30 days | 0.333 | 0.362 | 0.452 | 0.500 | 0.658 | 0.736 |
|  | 45 days | 0.259 | 0.259 | 0.498 | 0.500 | 0.737 | 0.779 |
|  | 60 days | 0.246 | 0.246 | 0.498 | 0.500 | 0.751 | 0.782 |
| Deaths | 21 days | 0.233 | 0.286 | 0.466 | 0.500 | 0.750 | 0.809 |
|  | 30 days | 0.210 | 0.227 | 0.525 | 0.500 | 0.740 | 0.822 |
|  | 45 days | 0.194 | 0.207 | 0.528 | 0.500 | 0.787 | 0.854 |
|  | 60 days | 0.396 | 0.396 | 0.396 | 0.500 | 0.592 | 0.715 |

**eFigure 1:** For the secondary analysis defining early and delayed with respect to lifting stay-at-home (SAH), point estimates (95% confidence intervals) for the expected relative growth of COVID-19 cases (A) and deaths (B) under early (blue) and delayed (red) implementation of the public masking mandate over time.

(A)

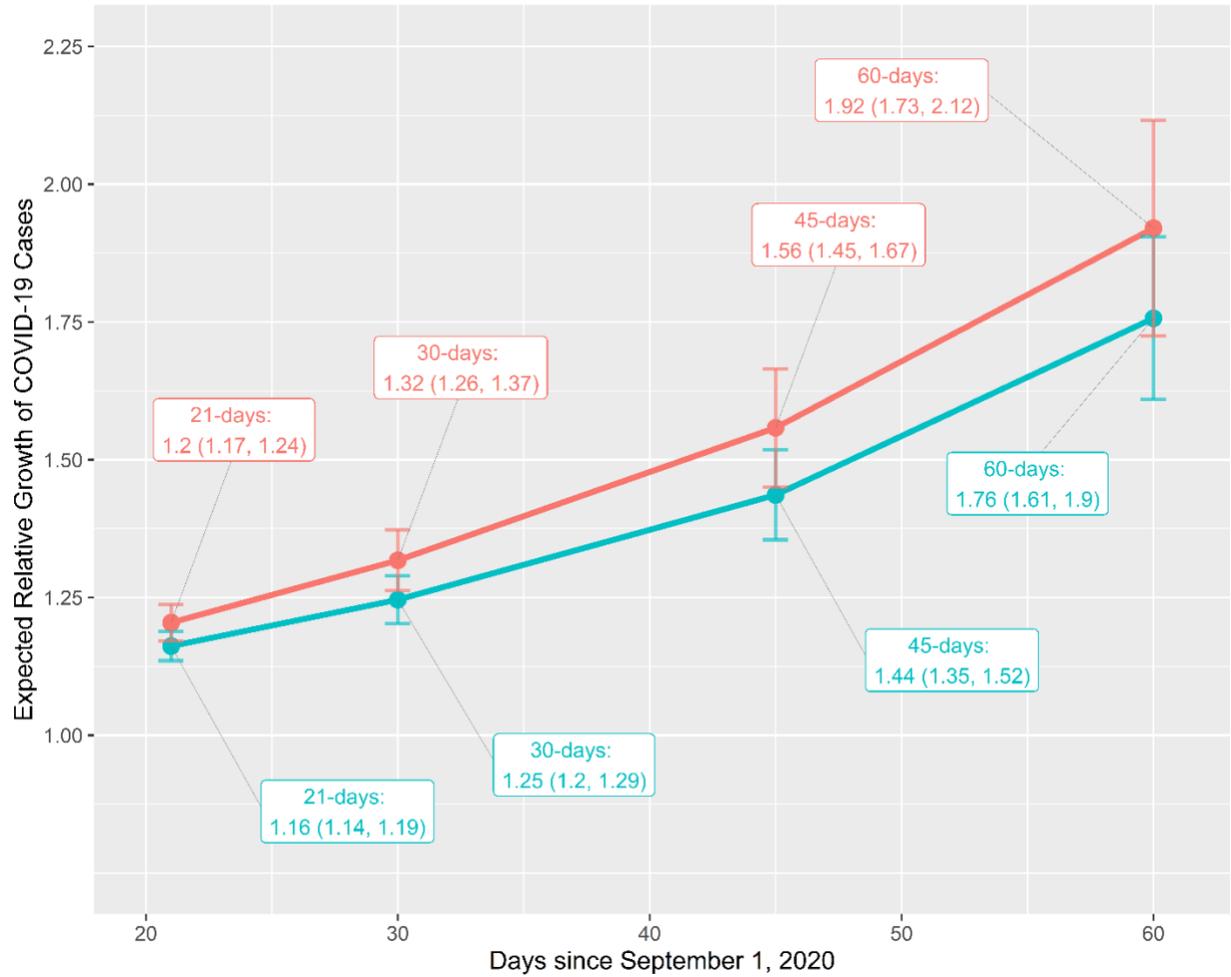

(B)

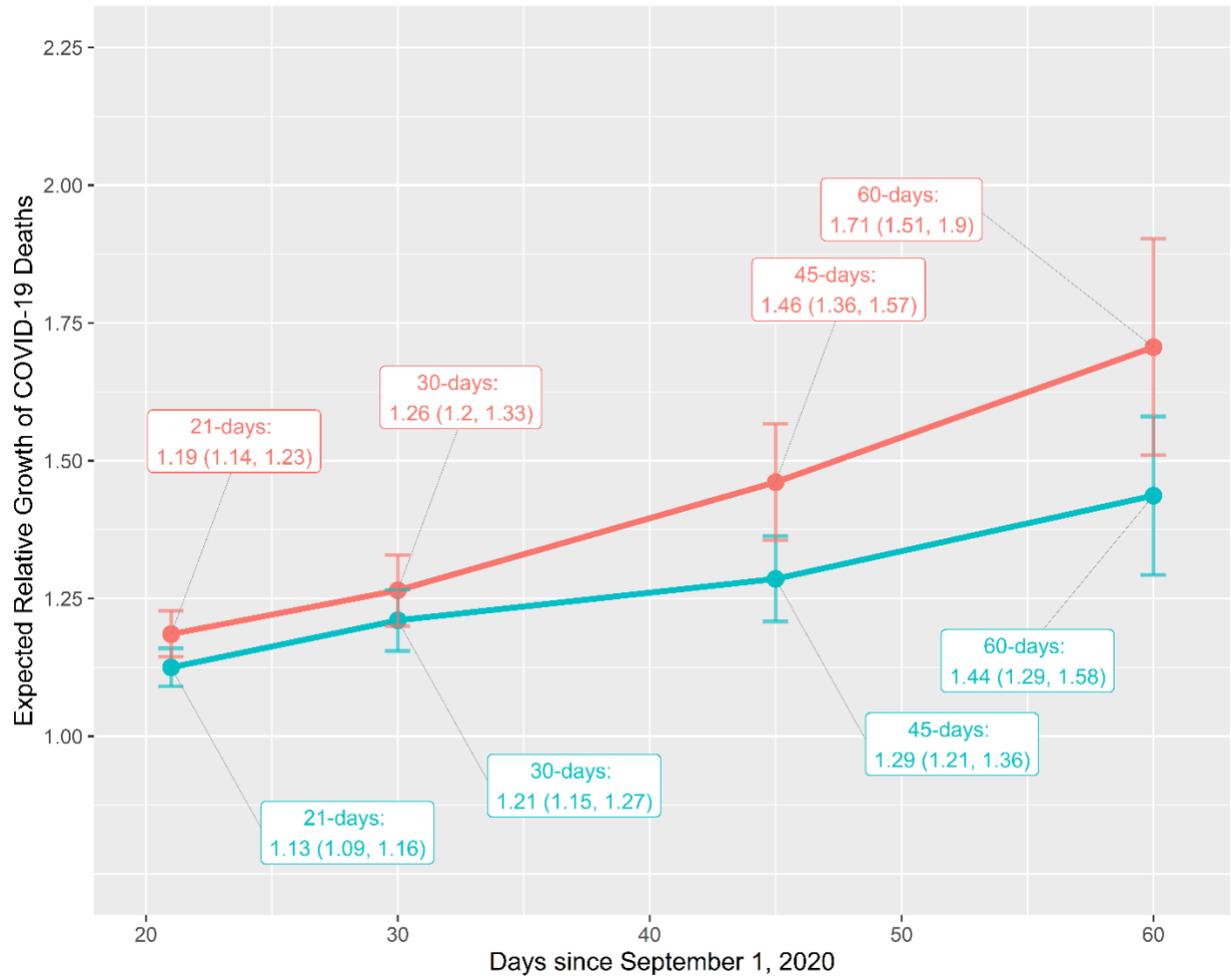

**eFigure 2:** For the secondary analysis defining early and delayed with respect to lifting stay-at-home (SAH), point estimates (95% confidence intervals) for the expected relative growth of COVID-19 cases (A) and deaths (B) under early (blue) and delayed (red) implementation of the public masking mandate and their ratios (C & D) over time.

(A)

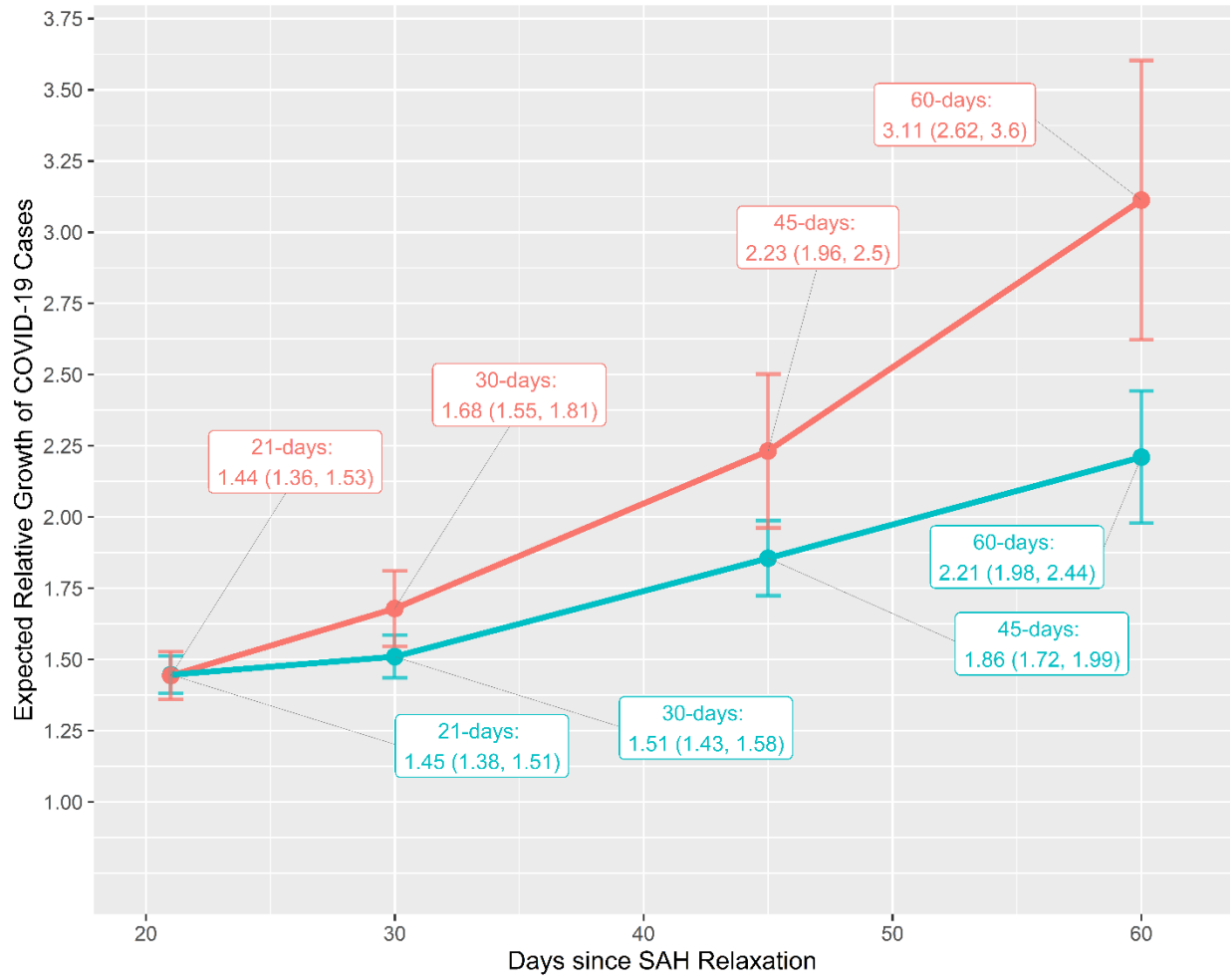

(B)

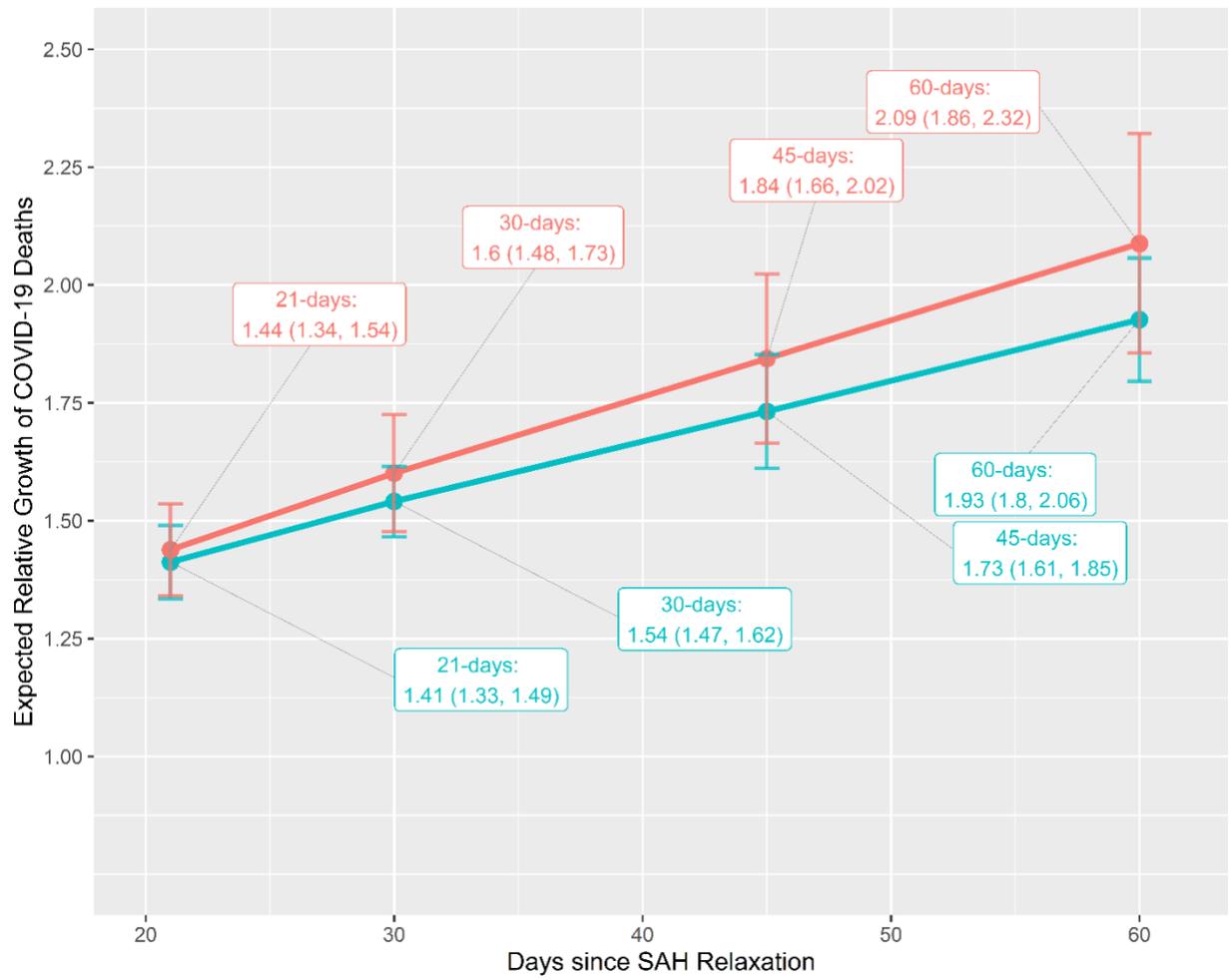

(C)

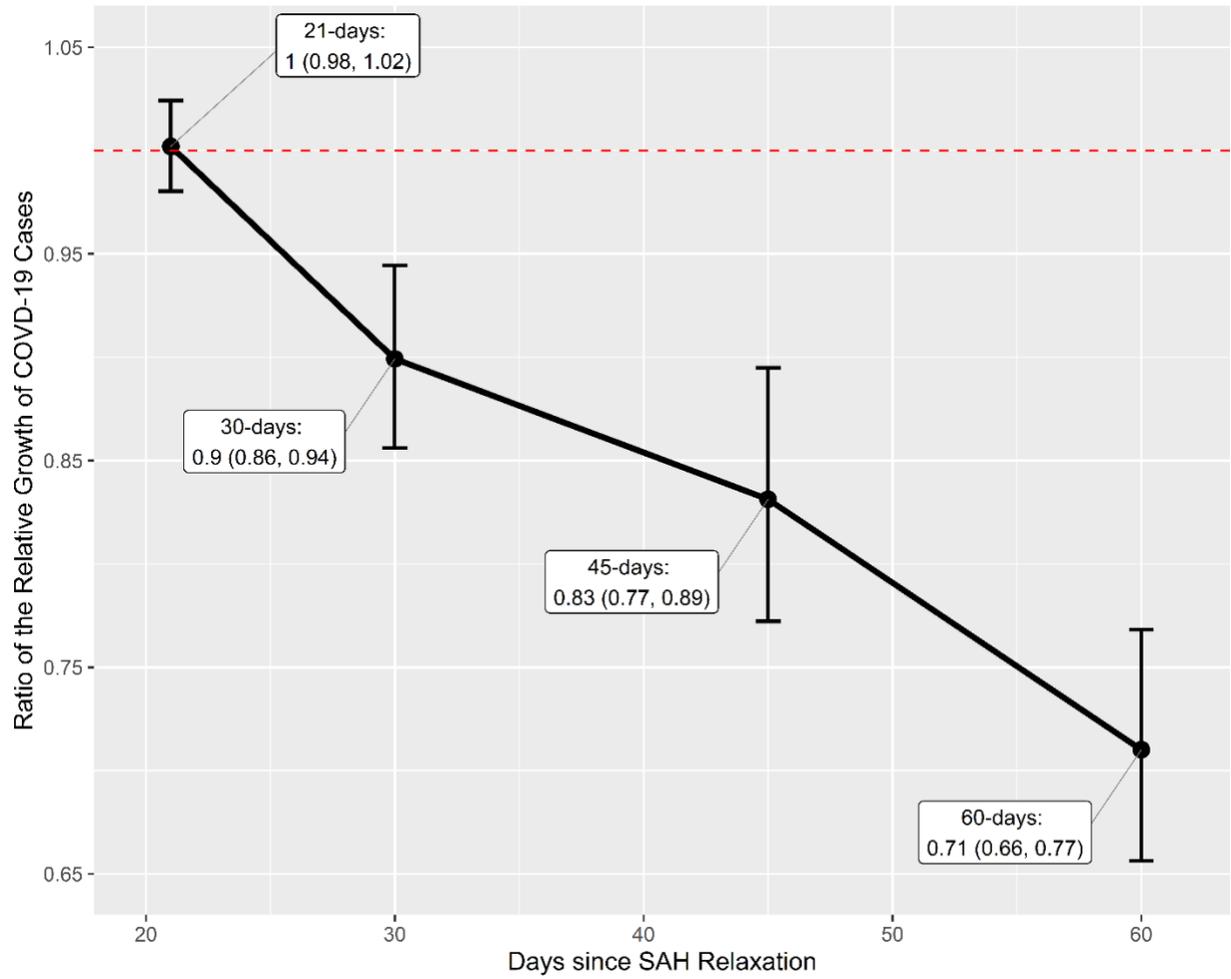

(D)

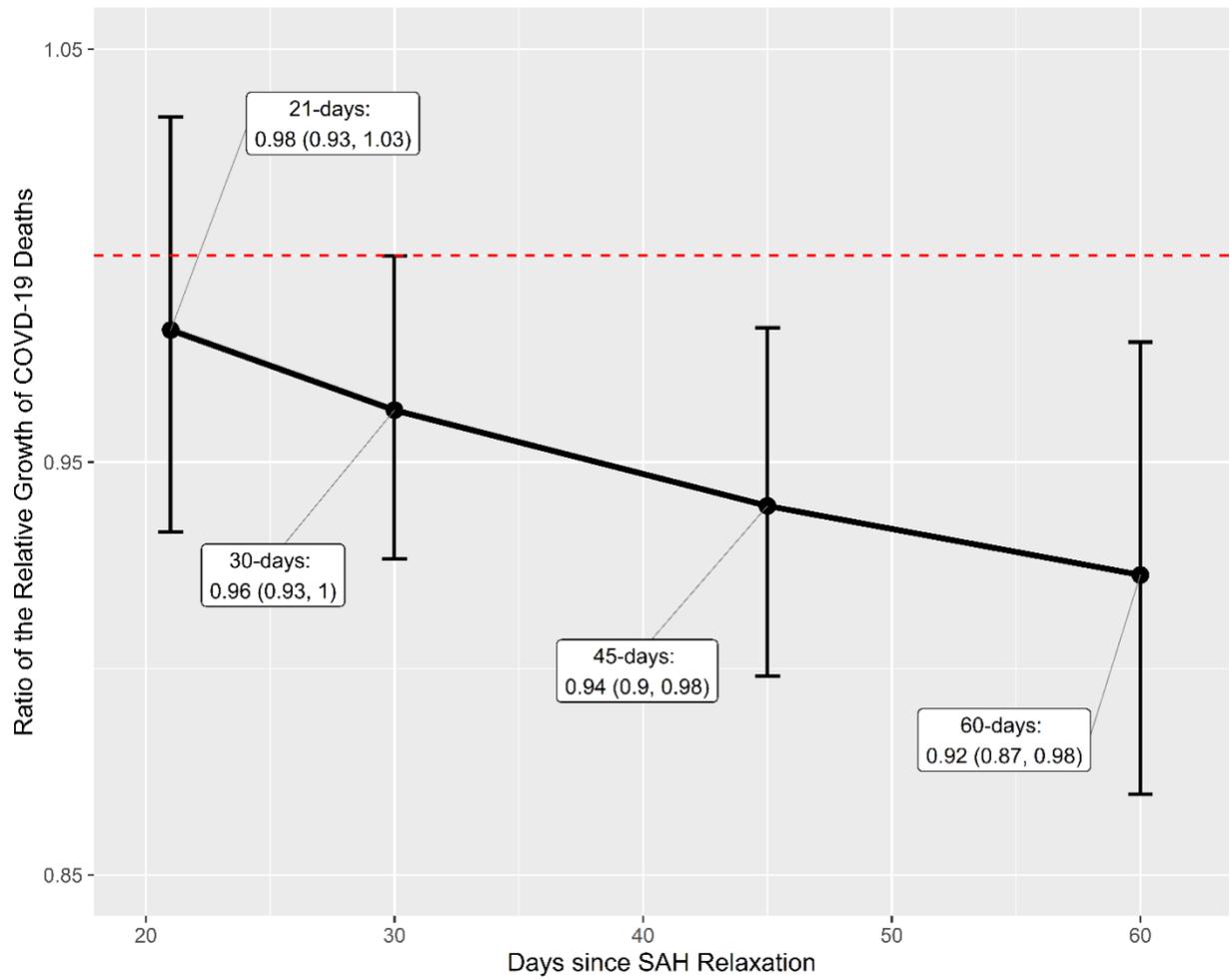